\documentclass[useAMS,usenatbib]{mnras}

\usepackage{amsmath, amssymb}
\usepackage{wasysym}
\usepackage{graphicx, txfonts, natbib,color}
\usepackage{subfigure}
\usepackage{epsf}
\usepackage{epstopdf}
\usepackage{longtable,lscape}
\usepackage{morefloats}
\usepackage{multirow} 
\usepackage[normalem]{ulem} %allows strikethrough text!
\newcommand{\eg}[0]{$\textnormal{e.g. }$}
\newcommand{\ie}[0]{$\textnormal{i.e. }$}
\newcommand{\Msun}[0]{\,\textnormal{M}_{\textnormal{\astrosun}}}
\newcommand{\asun}[0]{_{\textnormal{\astrosun}}}
\newcommand{\Zsun}[0]{\,\textnormal{Z}_{\textnormal{\astrosun}}}

\newcommand{\tn}[1]{\textnormal{#1}}

\newcommand{\blue}[0]{\color{blue}}

\newcommand{\msun}{\mbox{M$_{\odot}$\,}}

\newcommand{\zsun}{\mbox{Z$_{\odot}$\,}}

\DeclareMathAlphabet{\mathsc}{OT1}{cmr}{m}{sc}
\def\testbx{bx}%
\DeclareRobustCommand{\ion}[2]{%
\relax\ifmmode
\ifx\testbx\f@series
{\mathbf{#1\,\mathsc{#2}}}\else
{\mathrm{#1\,\mathsc{#2}}}\fi
\else\textup{#1\,{\mdseries\textsc{#2}}}%
\fi}
\newcommand{\ha} {\mbox{H$\alpha$}\,}

\newcommand{\Ha} {\mbox{H$\alpha$}\,}

\newcommand{\Nii} {[\ion{N}{ii}]\,}

\newcommand{\Oiii} {[\ion{O}{iii}]\,}

\def \mnras {MNRAS}
\def \apj {ApJ}
\def \apjl {ApJL}
\def \aap {A\&A}

\def \araa {ARAA}

\def \apjs {ApJS}

\title[Sub-solar metallicity SLSNe]{Superluminous supernova progenitors have a half-solar metallicity threshold}

\author[T.-W.~Chen et al.]{T.-W.~Chen$^{1,2}$\thanks{E-mail: jchen@mpe.mpg.de}, S. J.~Smartt$^{2}$, R. M.~Yates$^{1}$, M.~Nicholl$^{3}$, T.~Kr{\"u}hler$^{1}$, P.~Schady$^{1}$, \newauthor
M.~Dennefeld$^{4}$, C.~Inserra$^{2}$\\
$^1$Max-Planck-Institut f{\"u}r Extraterrestrische Physik, Giessenbachstra\ss e 1, 85748, Garching, Germany\\
$^2$Astrophysics Research Centre, School of Maths and Physics, Queen's University Belfast, Belfast BT7 1NN, UK\\
$^3$Harvard-Smithsonian Center for Astrophysics, 60 Garden Street, Cambridge, Massachusetts 02138, USA\\
$^4$Institut d'Astrophysique de Paris, CNRS, and Universite Pierre et Marie Curie, 98 bis Boulevard Arago, F-75014 Paris, France\\
}

\begin{document}
\date{Accepted 10 May 2017. Received 17 May 2016.}
\maketitle

\begin{abstract}
Host galaxy properties provide strong constraints on the stellar progenitors of superluminous supernovae.  
By comparing a sample of 19 low-redshift ($z < 0.3$) superluminous supernova hosts to galaxy populations in the local Universe, we show that sub-solar metallicities seem to be a requirement.
All superluminous supernovae in hosts with high measured gas-phase metallicities are found to explode at large galactocentric radii, indicating that the metallicity at the explosion site is likely lower than the integrated host value. 
We found that superluminous supernovae hosts do not always have star-formation rates higher than typical star-forming galaxies of the same mass. However, we confirm that high absolute specific star-formation rates are a feature of superluminous supernova host galaxies, but interpret this as simply a consequence of the anti-correlation between gas-phase metallicity and specific star-formation rate and the requirement of on-going star formation to produce young, massive stars greater than $\sim$10-20\,\msun. 
Based on our sample, we propose an upper limit of $\sim 0.5\,\Zsun$ for forming superluminous supernova progenitors (assuming an N2 metallicity diagnostic and a solar oxygen abundance of 8.69). Finally, we show that if magnetar
powering is the source of the extreme luminosity then
the required initial spins appear to be correlated with
metallicity of the host galaxy. This correlation needs further work, but if it holds it is a powerful link between the supernova parameters and nature of the progenitor population. 
\end{abstract}


\section{Introduction} \label{sec:Introduction}
The host galaxies of superluminous supernovae (SLSNe) Type I are generally faint dwarf galaxies \citep{Ne11} which tend to have low gas-phase metallicities \citep{St11,Ch13,Lu13}, and high specific star-formation rates ($\textnormal{sSFR} \equiv \textnormal{SFR}/M_{*}$) (\citealt{Le15}). There is currently a debate in the literature as to which of these two properties is the key requirement for SLSN-progenitor formation. Evidence for \textit{low-metallicity} being the primary driver comes from \citet{Ch13}, who found the host galaxy of SN~2010gx to have a very low  oxygen abundance of $0.06 \Zsun$ ($12 + \log({\rm O/H}) = 7.45 \pm 0.10$ on the $T_{\rm e}$ scale). The high quality host galaxy spectrum  enabled a detection of  the [O\,{\sc iii}] 4363\AA\ line, providing a reliable ``direct'' method  estimate of the oxygen abundance  for the first time for any such SLSN host. This is still the lowest 
metallicity for any supernova (or low-redshift gamma-ray-burst, GRB) host galaxy measured to date. It is known from theory and observation that low-metallicity environments lead to more compact, faster rotating 
massive stars with weaker stellar winds \citep[e.g.][]{Yo06,Mar07,Mok07,Hun08,Br11}. 
This may favour magnetar formation, which is a viable proposed central engine model for SLSNe as proposed by 
\cite{Kas10} and \cite{Woo10}. The application of this model shows good quantitative agreement with observations of 
SLSNe \citep[e.g.][]{Ins13,Ni13} and a SN associated with an ultra-long GRB \citep{Gr15}.
In the first of the large sample papers, \citet{Lu14} proposed that the hosts of SLSNe Type I share the same low metallicity (sub-solar abundance) preference as long-duration GRB (LGRB) host galaxies. In fact, SLSNe Type I could require even lower-metallicity environments than LGRBs (\citealt{Ch15}). That the host galaxies of SLSNe Type I are consistently fainter than those of LGRBs is supported by HST imaging \citep{An16}. 
\citet{Vr14} showed that interstellar medium (ISM) absorption features may also indicate different environments for SLSNe and GRBs. 

\citet{Le15} proposed that high sSFR is the primary driver to produce SLSNe, and they found that half of their SLSN hosts are extreme-emission-line galaxies (\ie galaxies exhibiting emission lines with EW $> 100$\r{A}; e.g. \citealt{Ca17}). They suggested that the progenitors of SLSNe constitute the first generation of stars to explode in a starburst, 
therefore being even younger and more massive, than the progenitors of LGRBs. In contrast, \citet{Lu15} have instead suggested that SLSN Type I progenitors are older and less massive stars than those of LGRBs. This is based on the locations of SLSNe Type I within their hosts, which show less of a preference for the brightest regions of the host galaxy when compared to LGRBs. 
Another potential argument is that some SLSN host galaxies have been observed with possible high metallicities, which questions the proposal that low metallicity is a key requirement. For example, one of the most metal-rich host galaxies of a SLSN Type I is MLS121104 \citep{Lu14}, with $12 + \log$ (O/H) = 8.8 ($R_{23}$ scale, \citealt{KK04}; hereafter KK04). However the SN location is clearly offset from the host centre, and further investigation is required to determine if this is indeed the host, or if the metallicity at the SN position is the same as that of the observed galaxy. Also, the metallicity estimate needs verification with the detection of the \Oiii $\lambda4363$\AA\ line. In this work, we found that the host galaxy of MLS121104 has $12 + \log$ (O/H) = $8.30\pm0.02$ using 
the N2 scale of \citeauthor{PP04} (2004, hereafter PP04).

We note here that sSFR itself cannot be a direct physical cause of SLSN progenitor formation, as a star-forming region cannot know about the total SFR or integrated star-formation history (\ie $M_{*}$) of the whole host galaxy. Therefore, for this interpretation to be valid, sSFR must instead be indicative of a more physical property. For example, a high sSFR could indicate a young stellar population, which could therefore contain massive stars capable of forming SLSNe (\eg as proposed by \citealt{Le15}). Alternatively, a top-heavy stellar initial mass function (IMF) has been proposed for star-bursting regions in ultra-compact dwarf galaxies \citep{Da12,Ma12}. This would allow an increased number of massive stars to form in these environments, hence causing the preference for SLSN to occur in high-sSFR galaxies. We would also note
that adjusting the IMF has no observational support from resolved studies of massive stars in Local Group or Local Universe galaxies, 
across a factors 5-10 in metallicity 
\citep[see the reviews and references therein of][]{2006ApJ...648..572E,2010ARA&A..48..339B,2011ApJ...741L..26F,2011ASPC..440...29M}.

In this work, we systematically compare the metallicity and star-formation properties of a sample of 19 SLSN Type I host galaxies against star-forming galaxies in the local Universe, in order to better determine which is the most important parameter for producing SLSN progenitors.

\section{Observational Data}\label{sec:Observational Data}
\subsection{SLSN host sample}\label{sec:SLSN hosts} 

We have compiled a sample of SLSNe Type I host data either in our possession (\citealt{Ch13, Ch15, Ch15PhD, Ch16}) or published in the literature \citep{Lu14,Le15} for all objects below $z < 0.3$.  
In this work, we have supplemented this sample with two additional host galaxy metallicity measurements taken from available late-time SN spectra (SN~2015bn, Gaia16apd), and upper limits on the host galaxy SFR for an additional two hosts (LSQ12dlf, SN~2013dg). This makes up a sample of 19 low-$z$ SLSN Type I host galaxies.
The SFR (from the \ha luminosity) and stellar mass have been corrected to the same IMF from \citet{Ch03}. All stellar masses are taken from \citet{Sc16}, and they are thus all derived in a consistent way (except for Gaia16apd, which we measured ourselves but using the same galaxy fitting templates as in \citealt{Sc16}).
We have propagated all errors to obtain an overall uncertainty for each property. 
We also measure a detection limit for the \ha line where there is no reported detection to estimate an upper limit for the SFR in these cases. 
Those host properties are summarised in Table\,\ref{tab:host_data}.
We only consider SLSN Type I hosts below $z = 0.3$, with the argument that there should be only minor stellar-mass and metallicity evolution between  $z = 0.1$ (the minimum redshift in our sample) and $z = 0.3$. 
The difference in the cosmic star formation rate density between  $z \simeq 0.1$ and $z \simeq 0.3$ using equation 15 in
\citet{MD14} is only $0.011\,\Msun\,\tn{yr}^{-1}\,\tn{Mpc}^{-3}$.  The evolution in the gas-phase metallicity of star-forming galaxies  
between $z \simeq 0.08$ and 0.29 is less than 0.1 dex at $\textnormal{log}(M_{*}/\Msun) \sim 9.0$ \citep{Za14}.

\begin{table*}
 \begin{minipage}{175mm}
  \centering
  \caption{The low-redshift ($z < 0.3$) SLSN Type I host galaxy properties. The SFR (from dust-corrected H$\alpha$ luminosity) and stellar mass are calculated assuming a Chabrier IMF. Note: all stellar mass are from \citet{Sc16}, except Gaia16apd$^{*}$, which we measured ourselves but using the same galaxy fitting templates as in \citet{Sc16}. References are reported below. }
\label{tab:host_data}
\begin{tabular}[t]{lccccc}
\hline
Name & redshift & N2 (PP04) & $\log$ Stellar mass & SFR & sSFR  \\
& & (12 + $\log$(O/H)) & (\msun) & (\msun\,yr$^{-1}$) & (Gyr$^{-1}$) \\
\hline
PTF10hgi & 0.098 & 8.38 (0.05)$^{f}$ & 7.58$^{+0.29}_{-0.31}$ & 0.04 (0.04)$^{f}$ & 1.05 (1.05)  \\ 
SN~2010kd & 0.101 & 8.07 (0.05)$^{f}$ & 7.30$^{+0.25}_{-0.29}$ & 0.07 (0.01)$^{f}$ &  3.51 (0.52) \\
Gaia16apd & 0.101 & 8.05 (0.04)$^{k}$ & 7.40$^{+0.90}_{-0.80}$$^{*}$ & 0.35 (0.01)$^{k}$ & 13.93 (1.74) \\
PTF12dam & 0.107 & 8.10 (0.02)$^{b}$ & 8.89$^{+0.15}_{-0.30}$ & 4.83 (0.09)$^{f}$ & 6.22 (0.16)  \\ 
SN~2015bn & 0.114 & 8.18 (0.02)$^{i}$ & 7.50$^{+0.38}_{-0.35}$ & 0.03 (0.00)$^{i}$ & 0.95 (0.05) \\
SN~1999as (SN location) & 0.127 & $<$ 8.29$^{f}$ & - & 0.04 (0.02)$^{f}$ & - \\
SN~2007bi & 0.128 & 8.20 (0.06)$^{f}$ & 7.92$^{+0.20}_{-0.21}$ & 0.01 (0.00)$^{g}$ & 0.12 (0.01) \\
SN~2011ke & 0.143 & 7.82 (0.11)$^{c}$ & 7.50$^{+0.20}_{-0.18}$ & 0.39 (0.01)$^{c}$ & 12.25 (0.45)  \\
SSS120810 & 0.156 & $<$ 8.23$^{f}$ & 7.42$^{+0.21}_{-0.17}$  & 0.06 (0.04)$^{f}$ & 2.28 (1.52)  \\
LSQ14an & 0.163 & 7.98 (0.04)$^{c}$ & 8.54$^{+0.13}_{-0.17}$ & 1.01 (0.02)$^{c}$ & 2.92 (0.07) \\
SN~2012il & 0.175 & 8.09 (0.02)$^{c}$ & 8.20$^{+0.18}_{-0.17}$ & 0.32 (0.01)$^{c}$ & 2.01 (0.08) \\
PTF11rks & 0.192 & 8.42 (0.15)$^{h}$ & 8.96$^{+0.12}_{-0.14}$ & 0.31 (0.03)$^{e}$ & 0.34 (0.03) \\
SN~2010gx & 0.23 & 7.97 (0.06)$^{a}$  & 7.97$^{+0.14}_{-0.13}$ & 0.41 (0.01)$^{a}$  & 4.42 (0.13) \\
SN~2011kf & 0.245 & $<$ 8.29$^{f}$  & 7.58$^{+0.19}_{-0.22}$ & 0.15 (0.05)$^{f}$  & 3.95 (1.32) \\
LSQ12dlf 	 &	0.250	& -    & 	7.56$^{+0.33}_{-0.34}$ & $<$ 0.004$^{j}$ & $<$ 0.11 \\		
LSQ14mo & 0.256 & 8.18 (0.05)$^{d}$ & 7.89$^{+0.15}_{-0.19}$  & 0.06 (0.01)$^{d}$ & 0.81 (0.13) \\
PTF09cnd & 0.258 & 8.24 (0.06)$^{f}$ & 7.87$^{+0.20}_{-0.21}$ & 0.21 (0.05)$^{f}$ & 2.83 (0.68) \\
SN~2013dg & 0.265 & - & 7.09$^{+0.82}_{-0.70}$ & $<$ 0.003$^{j}$ & $<$ 0.24 \\
MLS121104 & 0.303 & 8.30 (0.02)$^{h}$ &  9.27$^{+0.25}_{-0.24}$  & 0.79 (0.02)$^{e}$ & 0.43 (0.02) \\
\hline
\end{tabular}
\end{minipage}
\begin{flushleft}
\hspace{13em} a. \citet{Ch13} \\
\hspace{13em} b. \citet{Ch15} \\
\hspace{13em} c. \citet{Ch15PhD} \\
\hspace{13em} d. \citet{Ch16} \\
\hspace{13em} e. \citet{Lu14} \\
\hspace{13em} f. \citet{Le15} \\
\hspace{13em} g. \citet{You10} \\
\hspace{13em} h. adopted flux measurement from \citet{Lu14} \\
\hspace{13em} i. measured SN spectrum from \citet{Jer17}\\
\hspace{13em} j. measured SN spectrum from \citet{Ni14} \\
\hspace{13em} k. measured SN spectrum from Nicholl et al. in prep. \\
\end{flushleft}
\end{table*}

\subsection{Comparison galaxy sample}
\label{sec:11Mpc}
Unlike previous studies, which primarily focused on comparisons between SLSN hosts and GRB hosts, in this work we compare our SLSN sample with a Spitzer Local Volume Legacy (LVL) survey from \citet{Co14}, which consists of nearby galaxies within 11 Mpc, as well as the SDSS star-forming sample of \citet{Ya12}, which includes $\sim 110000$ galaxies with redshifts in the range $0.005 < z < 0.25$. Unlike the LVL sample, the SDSS sample is not complete, but its advantage is that it is at the same redshift regime as our SLSN host galaxy sample, and thus cosmic redshift evolution effects should not be a problem.
This allows us to better identify the key property required to produce SLSNe Type I \footnote{While this paper was in preparation, a paper using similar methods, but different sample was released as a preprint \citep{Pe16}. However we note that our work was carried out independently.}. 
We cross match the LVL sample with an H$\alpha$ imaging survey (\ie narrow-band photometry around the H$\alpha$ line) of galaxies within 11 Mpc sample taken from \citet{Ke08}. This leaves us with a final sample of 204 galaxies that includes both dwarf and giant irregular/spiral star-forming galaxies, spanning a wide luminosity range of $-9.6 < {\rm M}_{B} < -20.7$.

\citet{Ke08} give the H$\alpha$ luminosities for those galaxies, which we use to estimate the SFR. These luminosities are corrected for Milky Way reddening but not for internal extinction. Therefore, we have made our own internal dust extinction corrections using the Balmer decrement (H$\alpha$/H$\beta$) for 13 galaxies \citep{Mo10} within the 11 Mpc catalogue finding a correlation between their average $A_{V}$ and their observed  galaxy H$\alpha$ luminosities of $A_{V} = 0.9445 \times \log{\rm L}_{H\alpha} - 36.536$ (with a correlation of 0.70 and a $1\sigma$ scatter in $A_{V}$ of 0.6 mag). 
We use this relation to obtain extinctions for the remainder of the \citeauthor{Ke98} sample. Negative values of $A_{V}$ are set to zero. 
We then employ the calibration of \citet{Ke98} (which assumes an IMF of \citealt{Sa55}) to estimate galaxy SFR from the extinction-corrected ${\rm L}_{H\alpha}$, and apply a further correction to convert to a Chabrier IMF  by multiplying by a factor of 0.63.   

We take the stellar masses derived from {\em Spitzer} 3.6um measurements from \citet{Co14}, who assumed a mass-to-light ratio of 0.45 \citep{Mc15}. \citet{Mc15} found that their analysis is consistent with results based on galaxy SED fitting that assume a Kroupa or Chabrier IMF, and therefore a direct comparison of the \citet{Co14} stellar masses with our SLSN sample is justified.

\citet{Ke08} also provide the ratio [N \textsc{ii}]/H$\alpha$ from spectroscopic observations and from the correlation between [N \textsc{ii}]/H$\alpha$ and $M_{B}$, which we use to calculate metallicities via the N2 method of PP04. However, the ratio they provide is for [N \textsc{ii}] $\lambda\lambda$6548,6583, whereas the N2 scale uses only [N \textsc{ii}] $\lambda$6583. We estimated the ratio of [N \textsc{ii}] $\lambda$6583 and H$\alpha$ by applying the theoretical ratio between the [N \textsc{ii}] lines at 6548\r{A} and 6583\r{A} of 1:3, and then calculated their metallicities with the N2 scale.

\section{Results} \label{sec:Results}
\subsection{Metallicity versus sSFR}\label{sec:Z versus sSFR}

\begin{figure}
\centering
\includegraphics[width=0.5\textwidth]{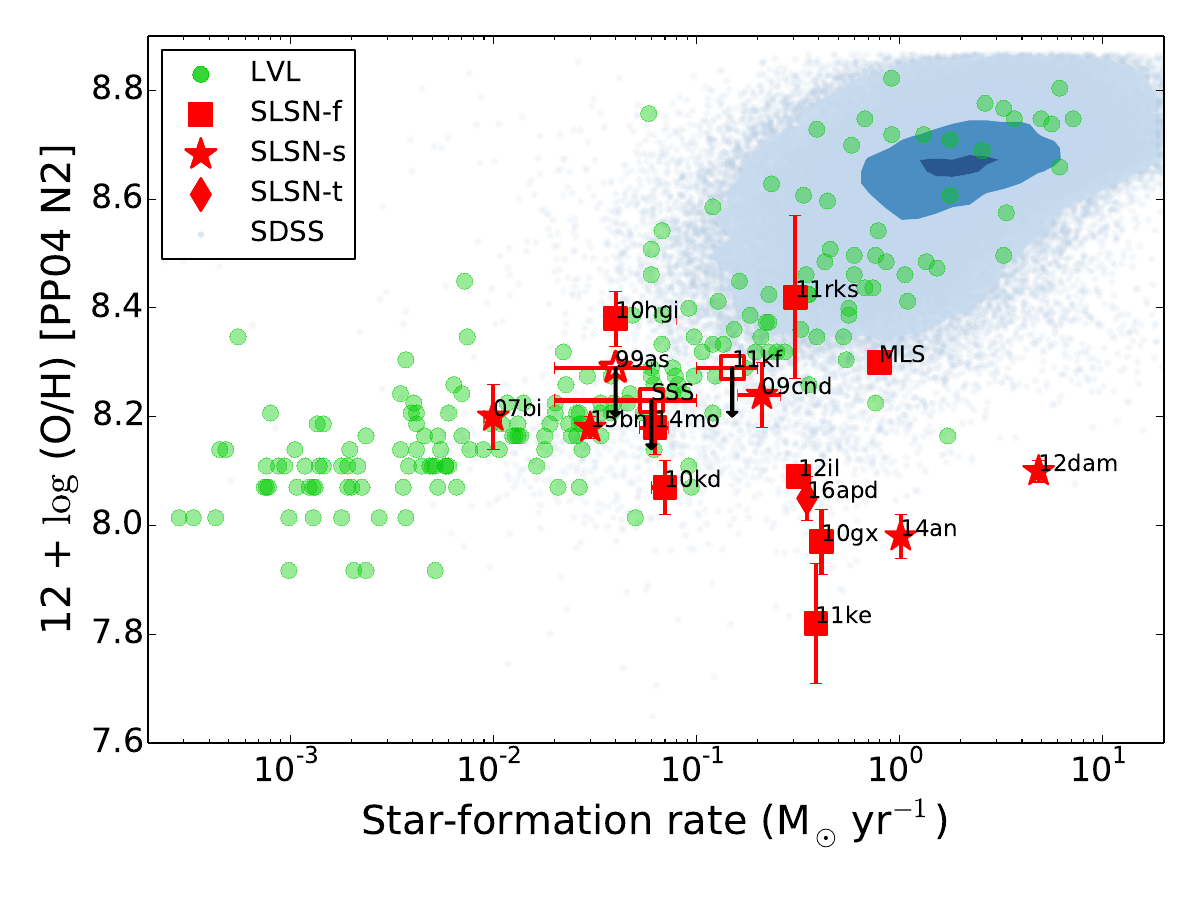}
\caption{Star-formation rate - metallicity relation for SLSN Type I host galaxies and the LVL and the SDSS galaxies.}
\label{fig:SFR-Z}
\end{figure}

Fig. \ref{fig:SFR-Z} shows the relationship between SFR and metallicity (N2 method) for 19 low-$z$ SLSN Type I hosts from the literature (red points, see section\,\ref{sec:SLSN hosts}). 
The LVL (green) and SDSS (blue) samples of nearby galaxies are also shown for comparison. SLSNe Type I seem to be divided into two groups based on light curve evolution. The majority have a fast rising and declining light curve (e.g. SN~2010gx), but a subset show a slow-evolving light curve (e.g. SNe~2007bi) \citep[see ][]{Ni15}. Recently the light curve of Gaia16apd was found to be an intermediate case between the fast- and slow-declining SLSNe \citep{Ka16}, hence we refer it as a transition object. We use different markers (square for fast-decliners, star for slow-decliners and diamond for transition) to highlight these three subclass of SLSNe, and we use these same markers throughout the paper.
We use the PP04 N2 metallicity diagnostic for all SLSN hosts, and for the LVL and SDSS samples so that we can make a fair comparison of their relative metallicities. The N2 method is reliable in the low-metallicity regime where our SLSN hosts are located.    
We can see that the most star-forming SLSN hosts have significantly lower metallicities than local galaxies with similarly high SFRs. No SLSN host in our sample has a measured metallicity higher than $12 + \log({\rm O/H}) \sim 8.4$, which is roughly half the solar oxygen abundance [assuming $12+\textnormal{log(O/H)}\asun = 8.69$, \citealt{A09}]. 
For comparison, \citet{{Mod11}} measured oxygen abundances at the SN position of 12 normal Type Ic SNe, and found that the metallicities of those hosts are all above 8.5 dex (PP04 O3N2 scale, which is similar to N2 scale).

\begin{figure}
\centering
\includegraphics[width=0.5\textwidth]{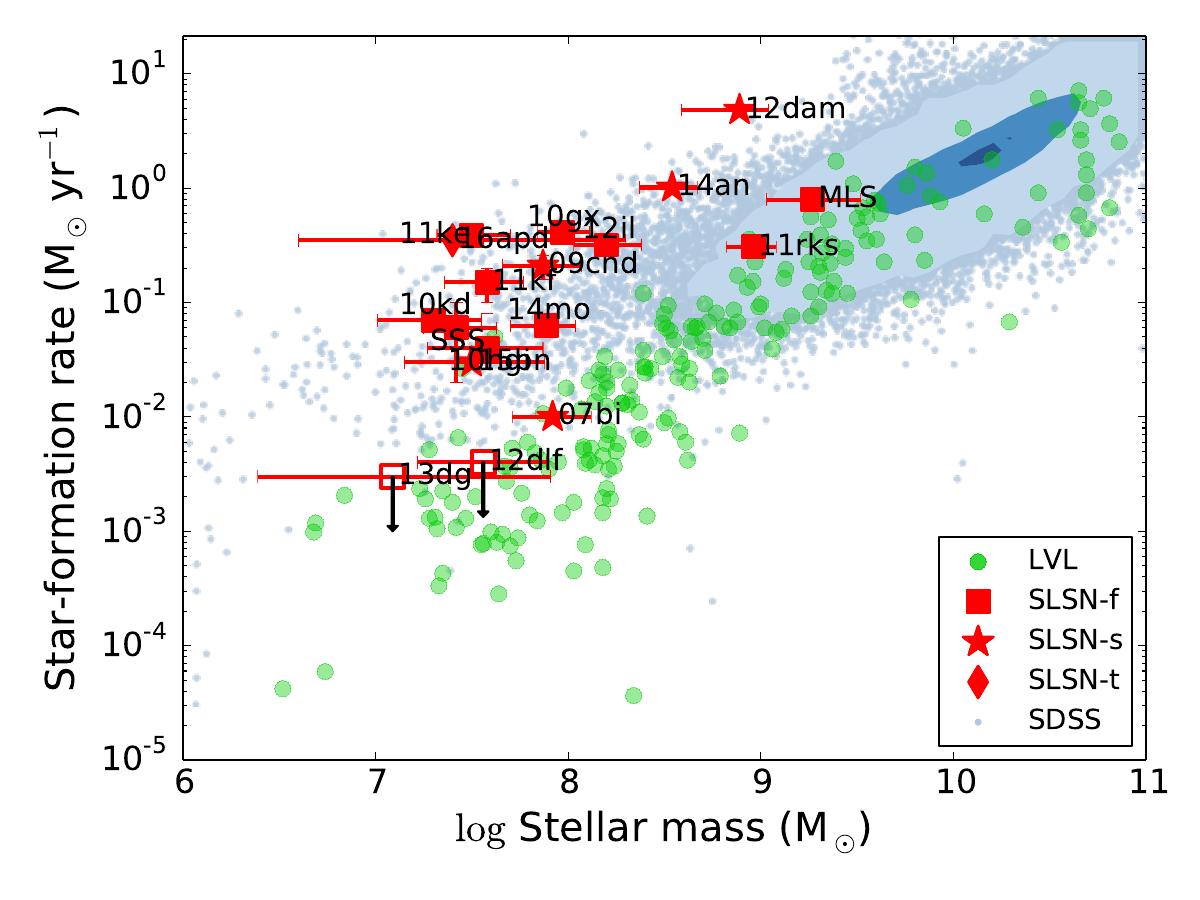}
\caption{The stellar mass - SFR relation for SLSN Type I host galaxies and the LVL and the SDSS galaxies.}
\label{fig:Lum-SFR}
\end{figure}

\begin{figure}
\centering
\includegraphics[width=0.5\textwidth]{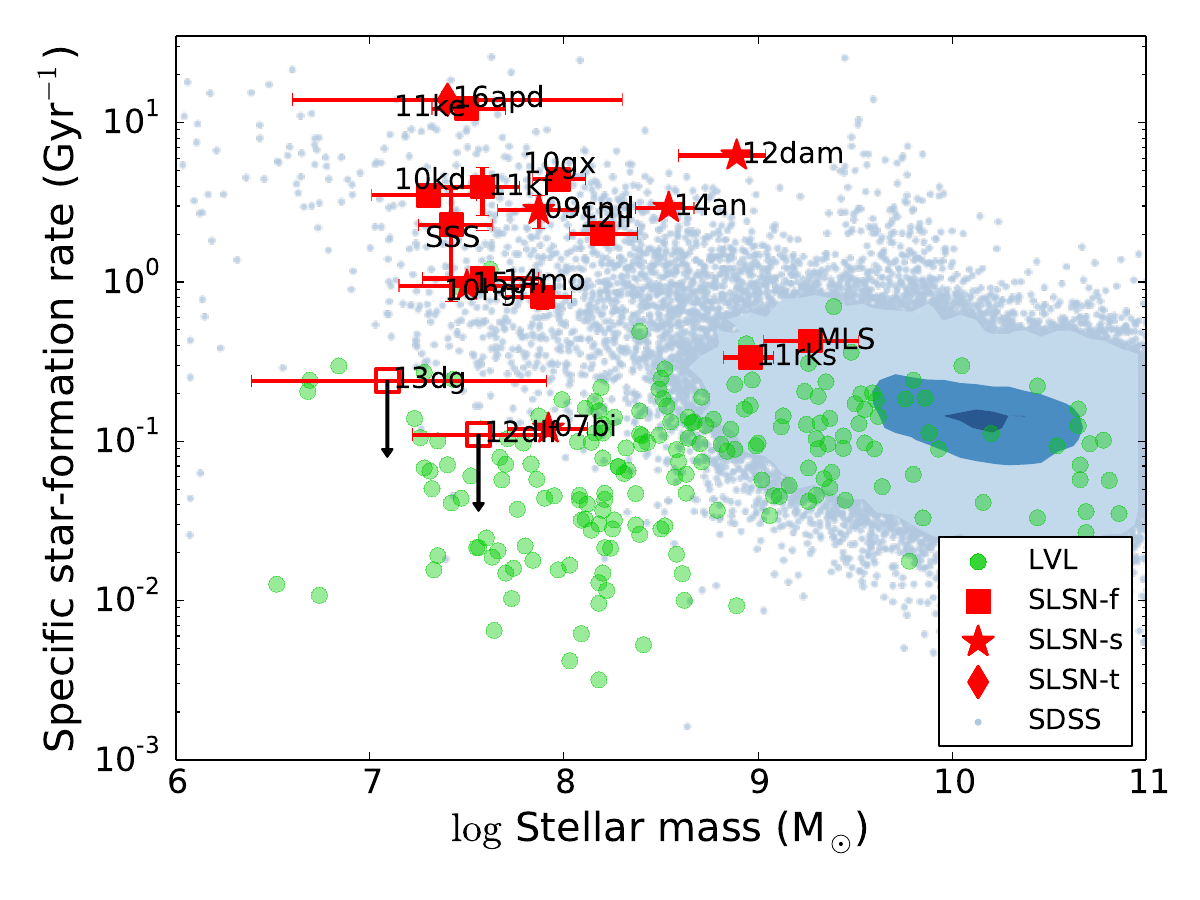}
\caption{The stellar mass - sSFR relation for SLSN Type I host galaxies and the LVL and the SDSS galaxies. SLSN hosts typically have high sSFRs compared to the overall local star-forming population.}
\label{fig:Lum-sSFR}
\end{figure}

Fig. \ref{fig:Lum-SFR} shows the relation between $M_{*}$ and SFR for the same {\blue three} samples. 
The LVL and SDSS samples diverge at the low stellar mass regime. SLSN hosts are more star-forming than LVL galaxies of the same stellar mass. However, they have typical (or slightly elevated) SFRs for their stellar mass compared to the SDSS sample, and lower SFRs than the bulk of the overall star-forming population.
Correspondingly, Fig. \ref{fig:Lum-sSFR} shows that most SLSN hosts have high sSFR in an absolute sense (\ie compared to the majority of the whole star-forming population in the SDSS). However, there are three hosts (SN~2007bi, LSQ12dlf and SN~2013dg) which exhibit lower specific SFRs than the rest of our SLSN-host sample. We note that the derived stellar mass values vary depending on the stellar population synthesis models assumed, even when the same IMF is used. For example, the stellar mass of SLSN hosts given in \citet{Sc16} is on average 0.2 dex lower than other literature \citep{Pe16} while comparing the same host galaxies, which introduces 2.5 times higher sSFR.

\begin{figure}
\centering
\includegraphics[width=0.5\textwidth]{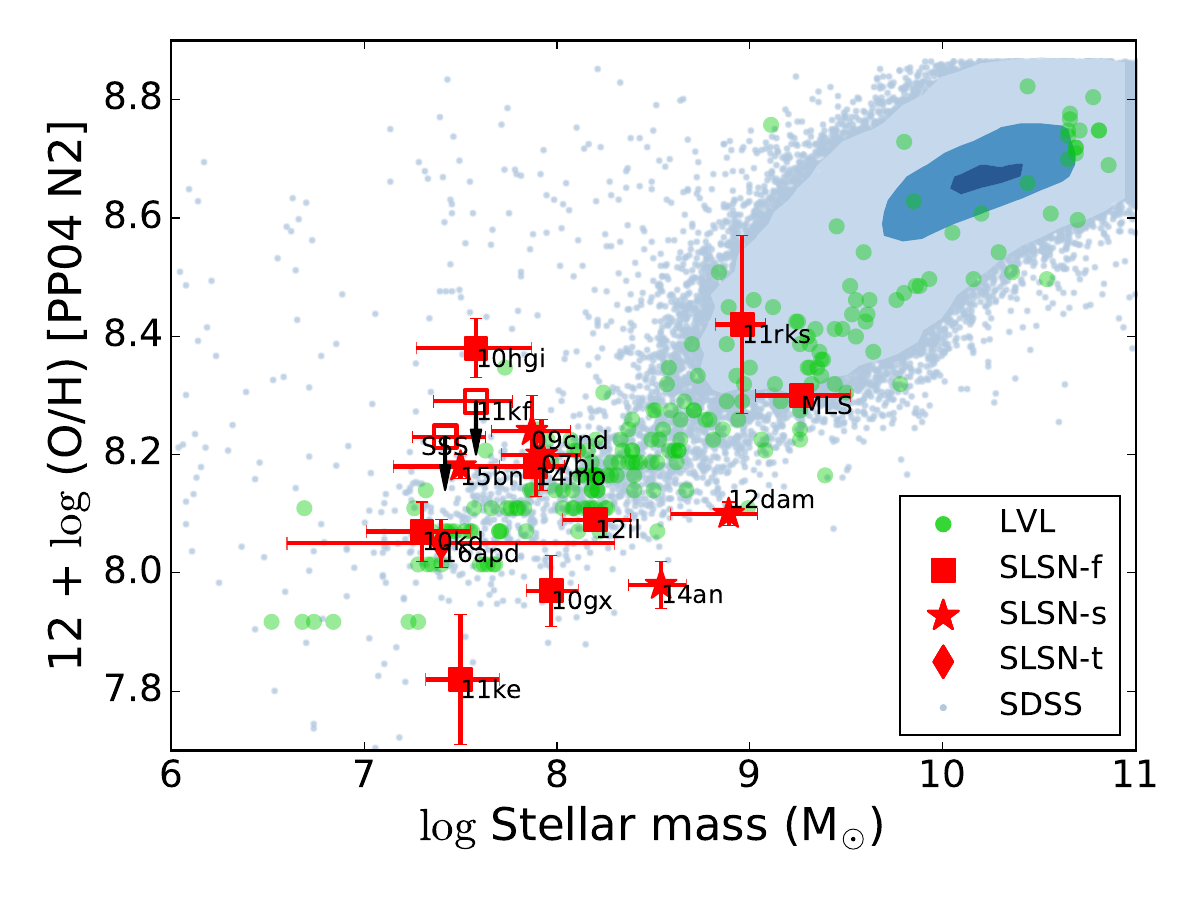}
\caption{The stellar mass - metallicity relation for SLSN Type I host galaxies and the LVL and the SDSS galaxies. No SLSN hosts above $12 + \log {\rm (O/H)} > 8.4$.}
\label{fig:mass_N2}
\end{figure}

Similarly, Fig. \ref{fig:mass_N2} shows that most SLSN hosts have comparable or lower metallicities than local galaxies of the same stellar mass. These systems are qualitatively consistent with the low-redshift fundamental metallicity relation (FMR, \citealt{Ma10}), which suggests an anti-correlation between SFR and metallicity at low $M_{*}$. However, two of our SLSN hosts, PTF10hgi and SN~2015bn (and possibly also PTF09cnd), have metallcities \textit{higher} than typical galaxies of the same mass (though still below 8.4, using the N2 method). This demonstrates that SLSN only require progenitors with a metallicity below some \textit{absolute} value, regardless of the typical metallicities found in galaxies of the same stellar mass.

\begin{figure}
\centering
\includegraphics[width=0.5\textwidth]{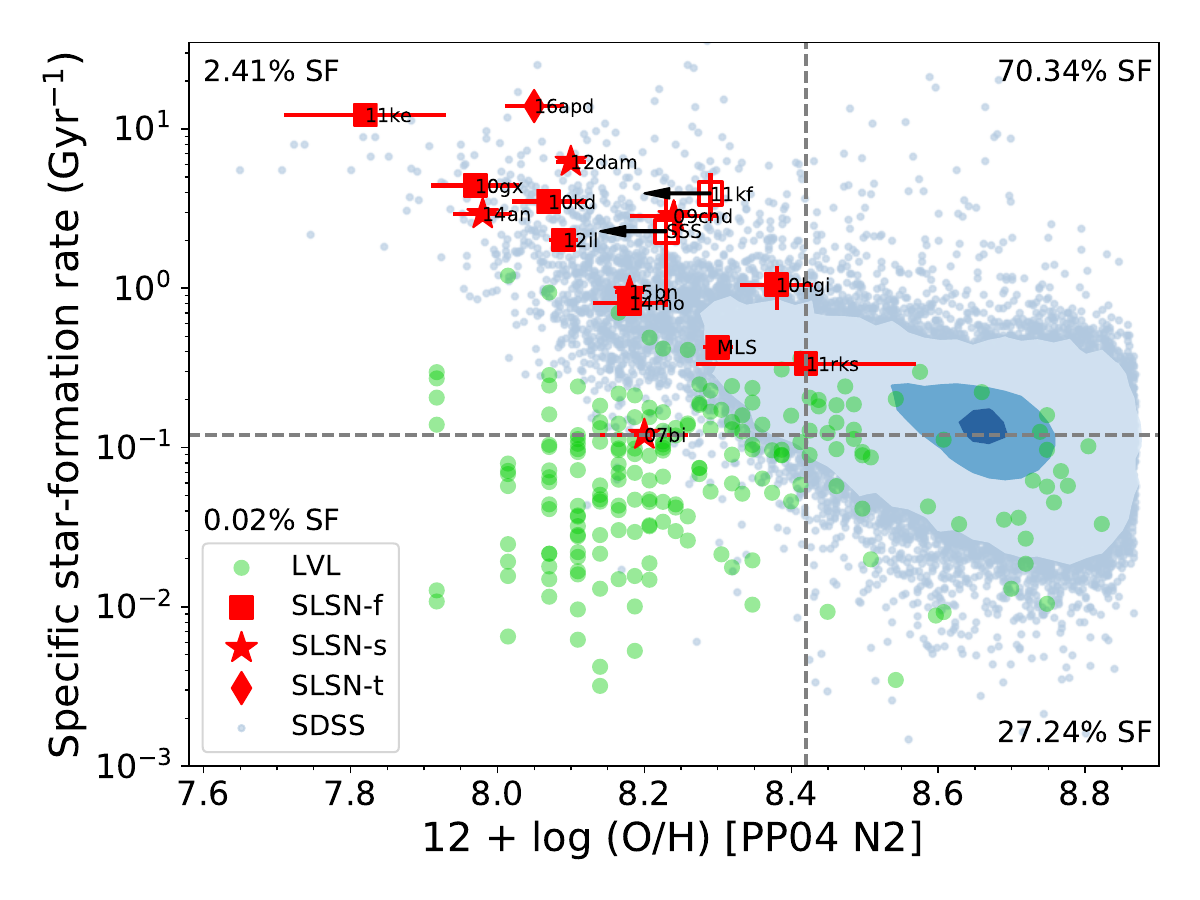}
\caption{The metallicity - sSFR relation for SLSN Type I host galaxies, compared to that of local irregular/spiral galaxies from the LVL and the SDSS samples. SLSN Type I hosts are clearly seen to reside in galaxies with relatively high sSFR and metallicities below 12 + $\log$(O/H) $\sim 8.4$.  The percentage of the total star-formation rate in the local galaxy sample that occurs in each of the quadrants marked by the dashed lines is provided.}
\label{fig:Z-sSFR}
\end{figure}

Turning our attention to the relative significance of low metallicity and high \textit{specific} star-formation rate, we present the metallicity-sSFR relation in Fig. \ref{fig:Z-sSFR}. This can be compared to the same relation presented by \citet{Pe16} in their fig. 11. We can see that our SLSN hosts lie in the lower-metallicity regime, and tend to have elevated sSFR compared to local star-forming galaxies. 
We note that the choice of metallicity diagnostic does not affect this conclusion. The parameter space in Fig. \ref{fig:Z-sSFR} has also been divided-up into four quadrants, such that our SLSN sample lies exclusively in the high-sSFR, low-metallicity quadrant. Only 2.41 per cent of the total star formation occurs in the quadrant associated with SLSN hosts. 

\begin{figure}
\centering
\includegraphics[width=0.5\textwidth]{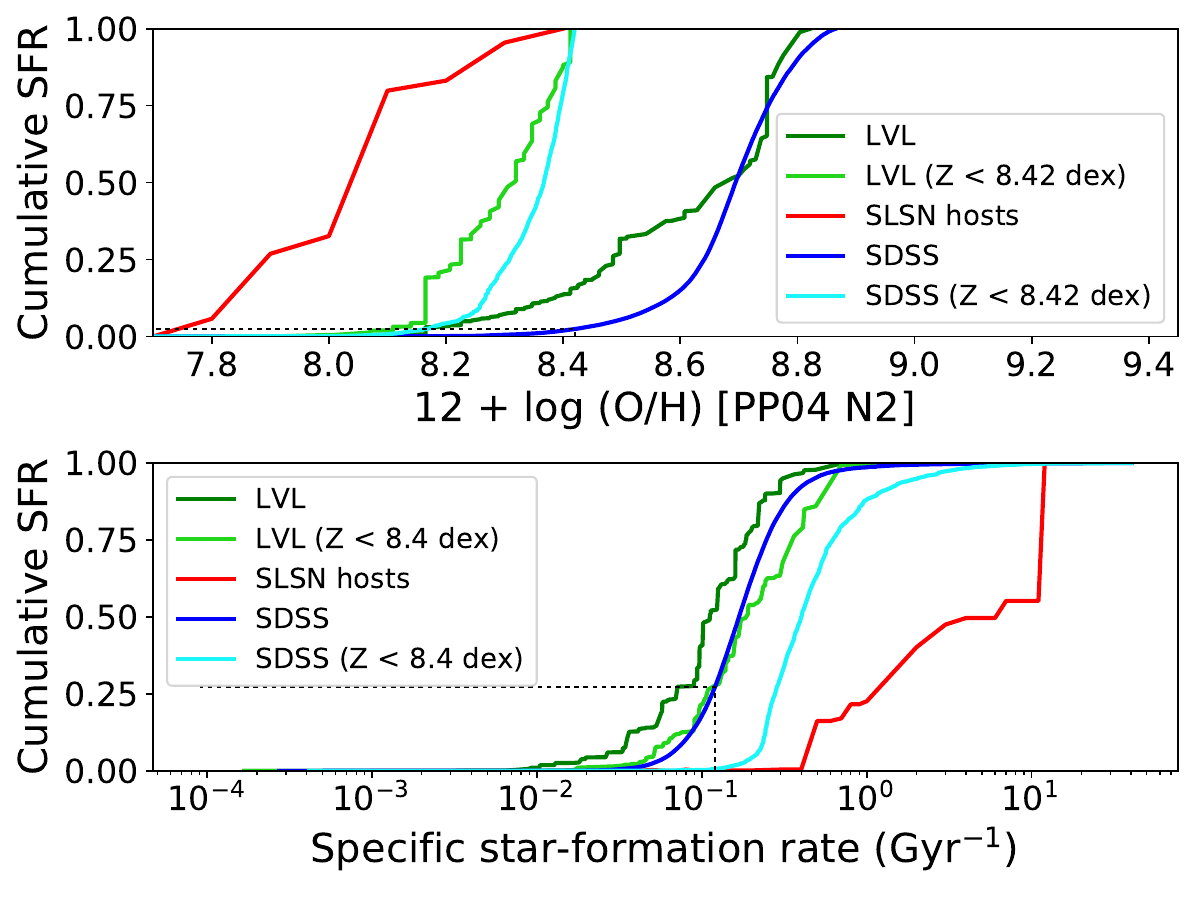}
\caption{Cumulative SFR distribution of the LVL and the SDSS galaxies as a function of N2 ratio (\ie [N \textsc{ii}]/H$\alpha$). About 2.41 per cent of star
formation occurs in the metallicity range where we observe SLSNe, suggesting that SLSN progenitors do not simply trace cosmic star formation.}
\label{fig:SFR_cum_dist}
\end{figure}

To investigate this further, in Fig. \ref{fig:SFR_cum_dist} we plot the cumulative distribution of SFR for SLSN hosts (red line) and the LVL sample (green colour lines) and the SDSS sample (blue colour lines) as a function of metallicity and sSFR. Only 7.85 per cent of the star formation in the local Universe occurs in galaxies with metallicities below 8.42 (\ie with metallicites similar to those of SLSN hosts). Even when only considering these low-metallicity systems, SLSNe still favour lower-metallicity star-forming regions than typically found in local galaxies. Similarly, SLSNe are preferentially found in the highest sSFR galaxies, even when only comparing to the low metallicity ($< 8.42$ dex) local population. 

These two results suggest that SLSNe do not simply trace star formation in the local Universe, but instead prefer both low-metallicity and high-sSFR environments. The probability (assuming a binomial distribution) that all of the 19 low-$z$ SLSN Type I hosts are in the low-metallicity bin simply by chance is only $(7113/295545)^{19} = 1.77 \times 10^{-31}$.

The metallicity-sSFR parameter space has also been studied for the sample of $\sim{}110,000$ emission-line galaxies from the SDSS-DR7 by \citet{YK14}. Such samples can be used to make qualitative statements about the type of galaxies which seem to host SLSNe. \citet{YK14} show that systems within the same range of sSFR and metallicity as our SLSN hosts have low stellar masses, high gas fractions, and young ages,  compared to typical star-forming galaxies in the local Universe (see their fig. 13). From analogy to the Munich semi-analytic model of galaxy evolution, \textsc{L-Galaxies}, they also show that these systems' gas-phase metallicities should be lower than their stellar metallicities (see panel E of fig. 3), indicating that recent accretion of low-metallicity gas is fuelling current star formation.

\subsection{Integrated metallicity versus SLSN offset}\label{sec:Z versus offset}
\begin{figure}
\centering
\includegraphics[width=0.5\textwidth]{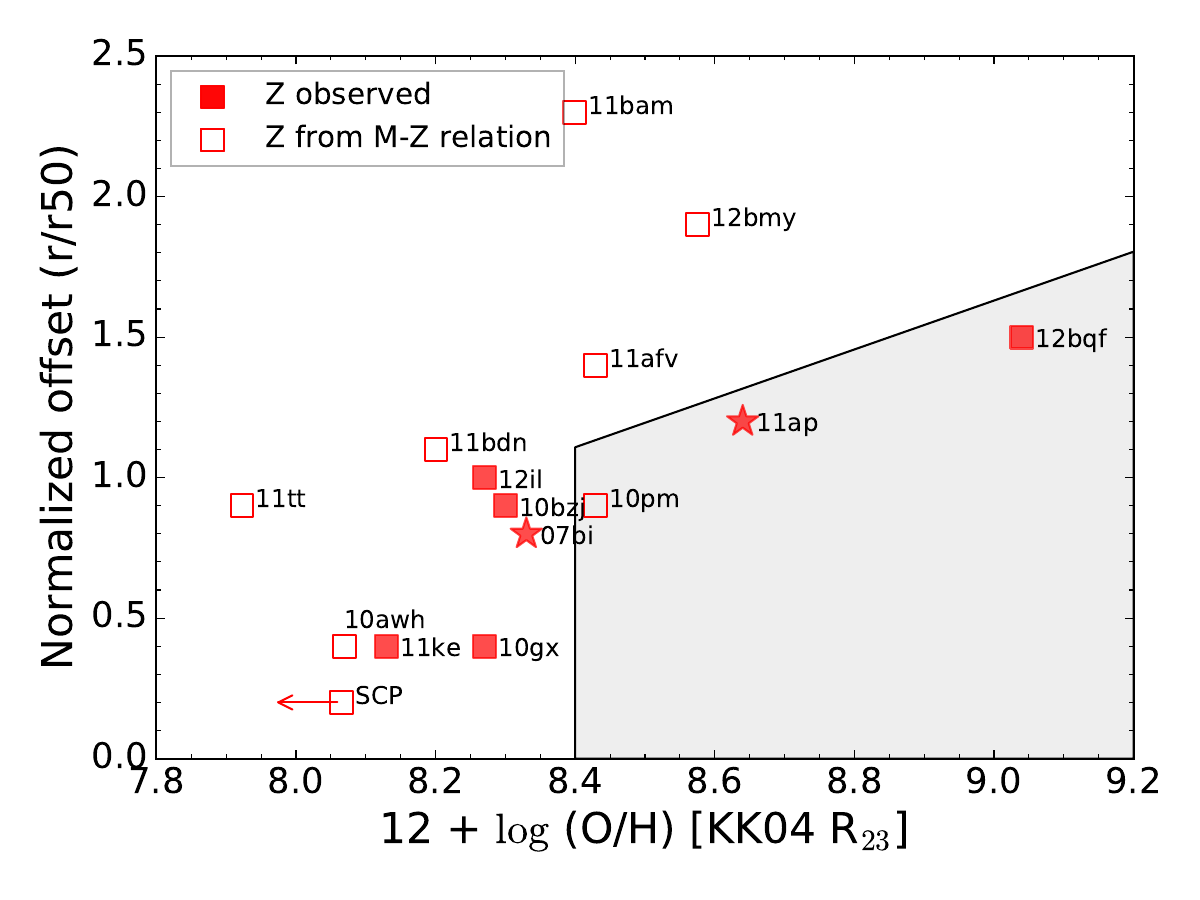}
\caption{The metallicity - offset relation for SLSN Type I host galaxies from the \citet{Lu14,Lu15} sample. Hosts with known metallicities (filled red symbols) and with metallicities predicted from the mass-metallicity relation (empty red symbols) are shown. There is a correlation between the reported host metallicity (here measured with the KK04 $R_{23}$ method) and the galactocentric distance of the SLSN. Crucially, there are no systems in the bottom-right region of the plot (grey area; bound by the best-fit relation and a half-solar metallicity threshold), indicating that no SLSN in our sample have exploded near the centres of high-metallicity galaxies.}
\label{fig:Z-radius}
\end{figure}

Further evidence for the preference of SLSNe to occur in low-metallicity environments is found in Fig. \ref{fig:Z-radius}, where we plot metallicity versus galactocentric distance of the SLSN event for the sample of \citet{Lu15}, who measured the SN position normalised to the rest-frame UV half-light radius ($r_{50}$)  from \textit{HST} images. 
This sample comprises 7 host galaxies with metallicity measurements  in the redshift range $0.12 < z < 0.65$, and 8 host galaxies  with only mass measurements in the redshift range  $0.74 < z < 1.6$. 
To obtain metallicities for these higher-redshift systems, we use a fit to the mass-metallicity relation for the lower-redshift systems: $Z_{\textnormal{g}}(\textnormal{KK04}) = 0.37\,\textnormal{log}(M_{*}) + 5.32$ (with a correlation of 0.92 and a $1\sigma$ scatter in $Z_{\textnormal{g}}$ of 0.11 dex). 
To account for the evolution in the mass-metallicity relation with cosmic time, we apply a shift in metallicity of $-0.16$ dex for host galaxies at redshifts around $z = 0.78$ and $-0.26$ dex for those around $z = 1.4$, following the evolution measured by \citet{Za14} for galaxies of stellar mass $\sim 10^{9} \Msun$. 

Fig. \ref{fig:Z-radius} shows a clear correlation between the host metallicity (in this case, measured with the KK04 $R_{23}$ method, since the \Ha and \Nii lines are out of the observed wavelength range) and SLSN offset. More precisely, we can say that there are no SLSNe found near the centres of high-metallicity galaxies. Systems with disturbed morphologies may contain low-metallicity pockets of gas where SLSNe could form
or the true host galaxy may be kinematically distinct. 
For example, in the case of LSQ14mo, \citet{Ch16} found evidence for a possibly interacting system, with the SLSN lying in a smaller satellite galaxy that had a 0.4 dex lower metallicity than that of the main galaxy system. It is possible that an interacting galaxy system could induce gas flows triggering star formation in low-metallicity regions.
\cite{Sa15} showed that disk abundance gradients in spiral galaxies are common and universal when expressed in terms of effective radius, even down to absolute magnitudes of $M_g \simeq -18$. This may also be an indicator that abundance variations are in play in lower luminosity galaxies, and indicates that the metallicity at a SLSN site could be lower than the integrated measured value. Therefore, even in galaxies with high integrated metallicities (such as the host of PS1-12bqf at $z = 0.52$), a sub-solar metallicity could still be present at the offset SLSN site. 
One observed example is SN~1999as, which has a large offset from its host centre \citep[10.7 kpc;][]{Le15}. Although a high host metallicity of 12 + $\log$(O/H) = 8.56 (PP04 N2) is measured, the metallicity measured at the SN location is more than 0.3 dex lower (12 + $\log$(O/H) $<$ 8.29) \citep{Lu15}.

\section{Discussion} \label{sec:Discussion}
As mentioned in Section \ref{sec:Introduction}, there are two dominant interpretations of current SLSN-host data. The first is that low-metallicity is the main requirement for SLSN formation. The second is that high sSFR is instead the main requirement. From the data presented in Section \ref{sec:Results} we cannot unambiguously distinguish between these two interpretations. Indeed, Fig. \ref{fig:Z-sSFR} demonstrates that both low metallicity and high sSFR are present in all of the SLSN host galaxies sampled here. However, in this work, we suggest that low metallicity is a fundamental cause, whereas high sSFR is only a consequence of the low metallicity.

Observations show a clear preference for SLSNe to occur in low metallicity environments, which  implies  low stellar mass host galaxies through the mass metallicity relation. We  expect  that SLSNe also require massive star progenitors, and thus they are more likely to arise in galaxies with elevated SFRs. Combining these two factors gives rise to the observed high sSFR.  
 This interpretation is supported by the fact that there are many hosts that have weak (e.g. the host of SN~2007bi), or undetected H$\alpha$ lines. For example, we estimate the detection limit of \ha fluxes from the SLSN spectra, and find the host of LSQ12dlf to have a detection limit of SFR $<$ 0.004 \msun yr$^{-1}$, the host of SN~2013dg to have a detection limit of SFR $<$ 0.003 \msun yr$^{-1}$, and the host of LSQ14bdq to have a detection limit of SFR $<$ 0.05 \msun yr$^{-1}$. These low SFR objects are not shown in figures due to the lack of information of other host properties (e.g. unknown host metallicities and stellar masses).

\citet{Le15} argued that mass is the key driver and that SLSNe Type I are the very first stars to explode in a starburst, and thus the progenitors are very massive stars \citep[for specific object see ][]{Th15}.
Our data are not inconsistent with this interpretation, although we would point out that 
very high masses are not quantitatively constrained in either study.  Along with the very young stars that provide
the extreme emission line ratios, virtually all hosts have stellar populations up to a few tens of Myr old detected, as traced by the UV and optical continuum \citep{Lu14,Le15,Ch15}. 
The spatial resolutions that ground-based spectrographs probe at $z > 0.1$ are typically more than 2 kpc, so a 
precise determination of the nature of progenitor stars is not yet possible. Distinguishing between progenitors
of $\sim$10-20 \msun\ and greater than 50 \msun\ is not easily achieved with the data.
SLSNe are rare, and occur only at a rate of about 1 in 10,000 - 20,000 of the core-collapse population \citep{Qu11,Qu13,McC15}.
It may be that a combination of very high mass (as traced by high sSFR) and low metallicity is required. Or it could be that a high sSFR is just a consequence of star formation in low-metallicity environments.

Alternative scenarios that could produce low gas-phase metallicities in the SLSN region, without requiring low $M_{*}$, are (a) significant infall of pristine gas onto a previously higher-metallicity host galaxy, and (b) variations in metallicity within a host itself (e.g. low-metallicity pockets or global metallicity gradients). Crucially, both these scenarios are consistent with the presence of a low integrated sSFR. Therefore, with a larger sample of SLSN host galaxies with measured sSFRs, we could  break the degeneracy discussed above and confirm our interpretation that low-metallicity is the key driver. 

\subsection{A possible link between progenitor metallicity and magnetar spin}
The identification of metallicity as the key parameter for forming SLSNe Type I has important implications for determining the power source underlying the explosion itself. As discussed in Section 1, the magnetar model is one of the preferred scenarios for producing SLSNe. Any link between galaxy environment and parameters of the powering mechanism would help elucidate not only 
progenitor star properties (e.g. mass or metallicity) but also the underlying mechanism producing SLSNe. 

\begin{figure}
\centering
\includegraphics[width=0.5\textwidth]{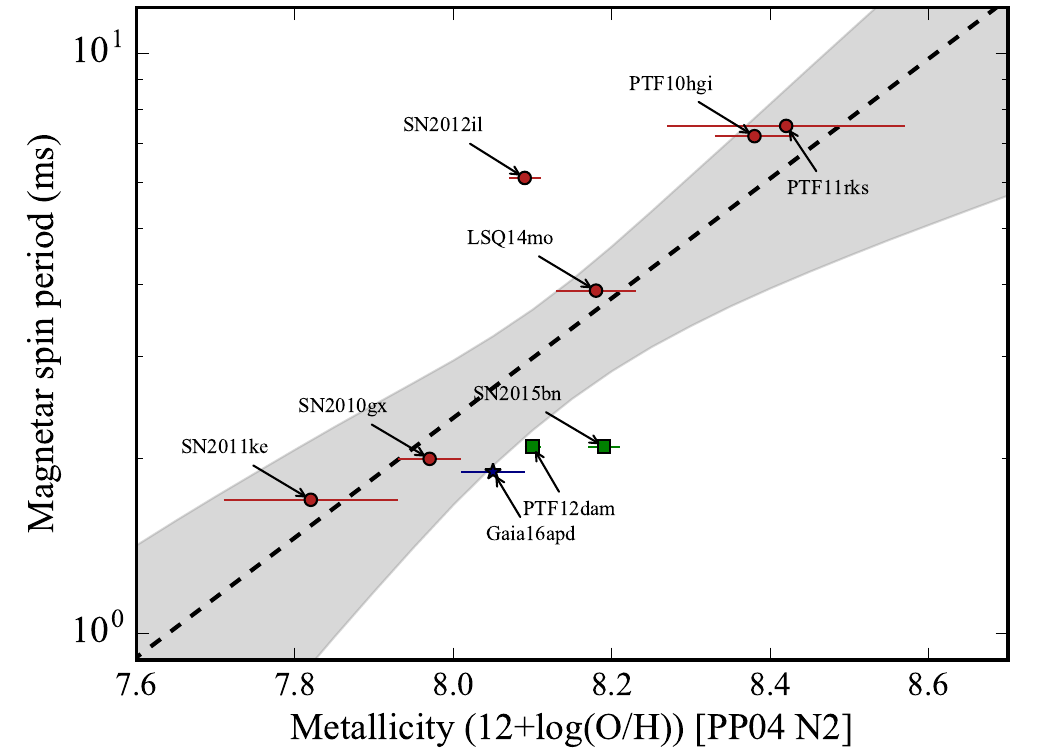}
\caption{The host metallicity - magnetar spin-period relation. The red markers show the fast-decliners and the green markers show the slowly-declining Type I SLSNe. The best fit shown by black dashed line and the grey area in which 68\% of all iteration are located.}
\label{fig:Z-spin}
\end{figure}

We show a possible correlation between host metallicity and magnetar spin-down period  in Fig.~\ref{fig:Z-spin} using the available data for eight SLSN and their hosts.
The magnetar spin-down period is derived from the SLSN bolometric light curve and semi-analytic models for magnetar energy deposition
(\citealt{Ins13,Ni13,Ni15}). The two quantities are truly independent measurements. 
A simple Spearman rank correlation analysis returns a coefficient $\rho = 0.85$. This corresponds to a null-hypothesis probability, that the observed correlation is the result of a statistical fluctuation, of $p = 0.003$.
This is an unexpectedly clear correlation between two quite independent observational properties. 

However measurement uncertainties play an important role and we hence test the significance of the observed correlation using a sample of $10^5$ bootstrapped distributions. Here, we vary the host metallicity according to the measurement errors and the spin-down period ($P_{\rm ms}$) by an assumed systematic error of 0.2 dex. We fit a linear relation to each of the data sets. The median of the resulting posterior distribution is obtained at $\log(\textnormal{\textit{$P_{\rm ms}$}}) = -7.89 + 1.03 \times (12+\log(\mathrm{O/H}))$. This best fit, and the area in which 68 per cent of all iterations are located, is shown in Fig.~\ref{fig:Z-spin}. 
No dependence of spin-down period 
on host metallicity is observed in only $p = 0.045$ of all cases.  A positive correlation is also found when using the $T_{\textnormal{e}}$ and KK04 $R_{23}$ metallicity diagnostics, with errors consistent with the N2-metallicity based data, although the relations obtained are somewhat steeper and have a slightly larger scatter.

The relation between the magnetar spin-down period and the host metallicity indicates that faster rotating magnetars reside in more metal-poor environments. Theory predicts that low metallicity massive stars are more compact and have a reduced mass loss, which 
results in faster rotation \citep{Ek12,Sc15}
This may lead to  chemically-homogeneous evolution which allows massive stars to retain more angular moment and thus rotate faster, even during the Wolf Rayet stage \citep{Yo06,Br11}. Observationally, massive stars with SMC metallicity ($\sim 0.2$ \zsun) appear to rotate more rapidly than  those with solar metallicity \citep{Mar07,Hun08,Ra13} and the difference may also be visible 
at LMC metallicites. Further work needs to be done to determine if stellar structure models including 
metallicity-dependent rotation can account quantitatively for faster-spinning magnetic neutron stars after core-collapse. 
If the observed relation shown in Fig.\,\ref{fig:Z-spin} holds true, it would strongly support the magnetar scenario, as it is not clear how the alternative SLSN progenitor models could bring about an equivalent relation between host galaxy metallicity, and light curve shape and peak luminosity. 

We caution that this is a simplistic picture and the spin periods of 
low metallicity stars and those of neutron stars formed after collapse 
are almost certainly affected by other parameters. A dependence on 
mass is likely and a wide range of ejecta masses have been found 
\citep[e.g.][]{Ni15}.
Stellar Binarity and separation distributions will critically affect
the final rotation rates of massive stars \citep{Mi13} and hence 
the spin rates of compact stars. All of these effects are likely to smear
out any correlation, hence it is perhaps surprising that we find such
a trend. More data may uncover scatter in this plot, or further 
analytic work may uncover a covariance between parameters which 
produces the effect. An obvious question to pose is if there is a 
simple observational parameter that can be plotted that is primarily 
affected by the spin period (before the semi-analytic models produce P)
and does that show a relation. The three parameters of B, P and M 
determine the overall shape of the light curve and its luminosity and there
is no one single observable that serves as a direct proxy for P.

\begin{figure}
\centering
\includegraphics[width=0.5\textwidth]{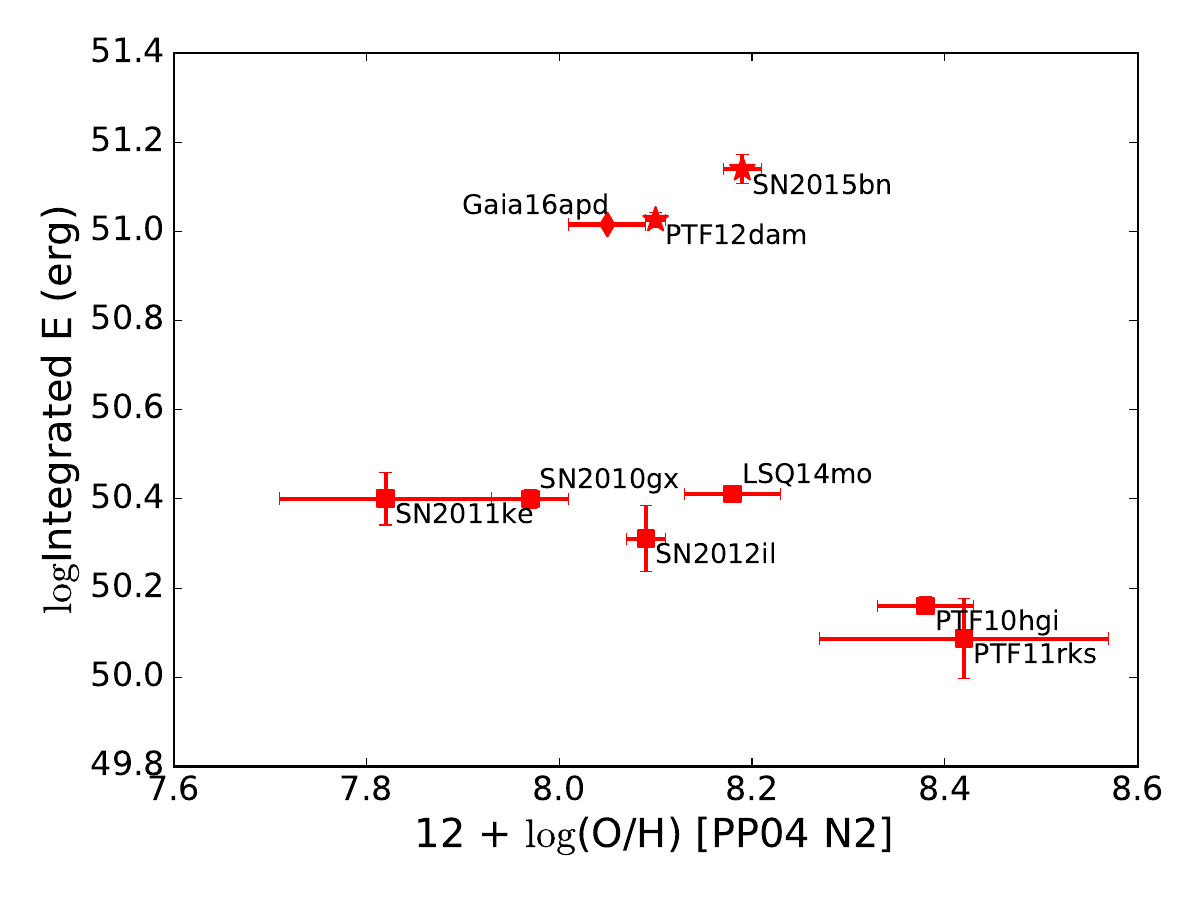}
\caption{The host metallicity - SLSNe integrated energy comparison. The fast dealing objects to sit on one locus and the slow declining objects to see above, in which the transition object also located. }
\label{fig:Z-eng}
\end{figure}

The only independent observational quantity that it makes sense to test in this 
way is the total integrated energy. Fig.\,\ref{fig:Z-eng} shows the integrated energy of SLSNe with their host metallicities, which provides a comparison between an observed (rather than model-dependent) SLSN property and the host galaxy. The energy is calculated by integrating a polynomial fit (order 3) of the bolometric light curve (in rest-frame) from Log($L_{max}$) / e, where Log($L_{max}$) is taken from the fit and $e$ is the neperian number \citep[similar method used in][]{Ni15}.
There is an interesting trend for the fast declining objects to sit on one locus and 
the slow declining objects to sit above. This is perhaps a trivial statement 
since the slow declining objects stay brighter for longer. 
However the integrated energy does not simply represent the spin 
period, since it is linked to B and to the mass through the spin down time 
and diffusion time \citep[see][]{Kas10}.

Pair-instability supernovae (PISNe) also require low metallicity \citep{La07,KWH11,Yu13}, in order to maintain a sufficiently massive helium core to explode in this manner. However, low metallicity is necessary (but not sufficient in itself) to interpret SLSNe as PISNe. Most SLSNe Type I show light curves  that  are clearly inconsistent with being PISN  \citep{Ins13}. The slowly-declining events, which have been claimed to be PISN candidates \citep{GY09,GY12}, seem to have similar hosts to the fast decaying events. 
They have been interpreted  as simply the high-mass counterparts of the fast-declining SLSNe \citep{Ni13,Ni15} and we find no distinction here in their host properties. The slowly declining objects are SN~2007bi, LSQ14an, PTF12dam, and SN~2015bn and these
do not appear significantly different to the rest of the sample in any plots.

\section*{Conclusions} \label{sec:Conclusions}
Comparing the metallicity, sSFR and SN offset for a sample of
19 SLSN Type I host galaxies we have found perhaps the strongest evidence yet that low metallicity is a key parameter for forming SLSNe Type I. 
If SLSNe simply followed SFR in the local Universe, we would expect to see many more at high metallicity, where the bulk of star formation occurs.  Instead, we find that SLSN trace regions of significantly lower metallicity and higher sSFR than those typically found in local galaxies. 
We suggest that the presence of high sSFR is a consequence of (a) the anti-correlation between gas-phase metallicity and sSFR and (b) the requirement of on-going star formation for massive stellar progenitors to form and to produce SLSNe.

We propose that current evidence supports a metallicity cutoff of about $0.5 \Zsun$, above which we do not find any SLSNe Type I in our sample. Low metallicity may favour a magnetar (central engine) model, in which reduced mass-loss helps to maintain high angular momentum at core-collapse. We find a surprisingly 
clear trend for the derived magnetar spin period (from supernova lightcurve fitting) to be correlated with metallicity. The SLSNe from lower metallicity galaxies require magnetars which have shorter spin periods. While this needs further investigation, the correlation supports both a low metallicity requirement for progenitor stars and
the model of magnetar powering.

\begin{table*}
 \begin{minipage}{175mm}
  \centering
  \caption{Magnetar model parameters for 9 SLSNe. We selecte SLSNe which have a full light curve coverage before and after the peak brightness and a good spectral coverage to apply for {\it K}-correction. We also choose the same magnetar code in order to have a consistent fitting result. }
\label{tab:magnetar_data}
\begin{tabular}[t]{lccccc}
\hline
Name & Type & P & B & M$_{\rm {ej}}$ & Reference  \\
& & (ms) & (14G) & (\msun) &  \\
\hline
SN~2011ke& fast & 1.7 & 6.4 & 8.6 & \citet{Ins13} \\
SN~2010gx & fast & 2.0 & 7.4 & 7.1 &\citet{Ins13} \\
Gaia16apd & transition & 1.9 & 2 & 4.8 & \citet{Ni17} \\
PTF12dam & slow &2.1&1.5&16& \citet{Ch15} \\
SN~2015bn & slow & 2.1 & 0.9 & 8.4 & \citet{Ni16} \\
LSQ14mo & fast &3.9 & 5.1&3.9& \citet{Ch16} \\
SN~2012il & fast & 6.1& 4.1 & 2.3 & \citet{Ins13} \\
PTF10hgi & fast & 7.2 & 3.6 & 3.9& \citet{Ins13} \\
PTF11rks & fast &7.5 & 6.8& 2.8 & \citet{Ins13} \\
\hline
\end{tabular}
\end{minipage}
\end{table*}

\section*{Acknowledgments} \label{sec:Acknowledgements}
T.-W.~Chen appreciates Alan Fitzsimmons for thesis advise, to Fabio Bresolin and Rolf Peter Kudritzki for host data collection; thanks to Sandra Savaglio for encouragement and to Dan Perley, Pavel Kroupa and Jeff Cooke for helpful discussions; thanks the organisers and participants of STScI workshop ``The  Mysterious  Connection  Between  SLSNe and GRBs'' for stimulating discussions. 
The research leading to these results has received funding from the European Research Council under the European Union's Seventh Framework Programme (FP7/2007-2013)/ERC Grant agreement n$^{\rm o}$ [291222] (PI : S. J.~Smartt). 
T.-W.~Chen, RMY and TK acknowledge the support through the Sofia Kovalevskaja Award to P. Schady from the Alexander von Humboldt Foundation of Germany.

\end{document}